\pdfoutput=1
\documentclass[11pt]{article}
\usepackage[preprint]{acl}
\usepackage{times}
\usepackage{latexsym}
\usepackage[T1]{fontenc}
\usepackage[utf8]{inputenc}
\usepackage{microtype}
\usepackage{inconsolata}
\usepackage{graphicx}
\usepackage{booktabs}
\usepackage{xcolor}
\usepackage{amsmath}
\usepackage{multirow}
\usepackage{enumitem}
\input{numbers}

\newcommand{\qwen}{Qwen2.5-3B}
\newcommand{\llama}{Llama-3.2-3B}
\newcommand{\qwenthree}{Qwen3-4B}
\newcommand{\pp}{\,pp}
\newcommand{\IfMacro}[3]{\ifcsname qn@#1\endcsname #2\else #3\fi}

\title{Is 4-Bit Quantization More Than Noise to a Chat Model's Safety Behaviour?\\
Attributing Behavioural Shifts with Matched Perturbations}

\author{Ege Kaya \\ University of British Columbia}

\begin{document}
\maketitle
\begin{abstract}
Chat models on phones and laptops are usually 4-bit copies of the model that was evaluated, and audits of such copies disagree about what quantization does to refusal. A 4-bit build is one of many weight perturbations of the same size, so a change in its behaviour could come from its specific rounding error or from any perturbation that large. In a pre-registered study of three 3--4B chat models, with paired tests, blind human labels and an 8-bit null, we find two 4-bit shifts of opposite sign: with a compliant reply prefilled, Qwen2.5-3B continues into harmful content more often, and on held-out jailbreaks Llama-3.2-3B complies \N{C2.cells.llama.4bit-g64.absdiff} points less often. Matched perturbations separate the two. Shuffled and Gaussian versions of the error average close to zero, while scaling the real error, from its mirror image to twice its size, changes behaviour near-linearly. At half the real error, Llama's likelihood is almost unchanged, yet it already complies \N{C3.llamafree.e2-scale0.5.absdiff} points less often, and llama.cpp's Q4\_K\_M, at the same bits per weight, moves neither model. Refusal under 4-bit quantization depends on the specific rounding error, not on bits per weight or likelihood, so each build has to be compared with its original.
\end{abstract}

\section{Introduction}

Chat models that run on laptops and phones are usually not the models that were evaluated. They are 4-bit copies, in which each weight is rounded to one of sixteen values, often made by a third party with a single command. A 4-bit copy is about a quarter of the size of the original and nearly as capable, and capability is usually all that is checked. Whether it still refuses the same requests is taken on trust.

Existing work does not settle the question. Rounding error can be constructed adversarially, so that a model that is harmless at full precision misbehaves once quantized \citep{egashira2024exploiting, egashira2025gguf}. For ordinary quantization, some audits find that low-bit builds refuse less \citep{kadadekar2026quality}, while others find little change at 4 bits or find that some models refuse \emph{more} \citep{prasad2026temperature}. A study of sixteen models finds that quantization shrinks decision margins rather than adding random noise, and affects different kinds of decision to different degrees \citep{wu2026decisions}. Quantization can evidently change refusal, but the sign and size of the change vary from model to model, and there is no account of why.

We suspect a simple reason. A 4-bit build is a single perturbation of the weights, and many other perturbations have the same size. Refusal is concentrated in the first few tokens of a reply \citep{qi2024shallow} and mediated by a single direction in the activations \citep{arditi2024refusal}, so it may be sensitive to any perturbation of that size. The direction of the change would then be a property of the model, and audits would disagree because their models differ. Alternatively, the structure of the rounding error could matter. The two explanations can be told apart by perturbing the weights by as much as 4-bit rounding does, in a pattern that is not quantization, and checking whether refusal changes in the same way. \citet{gong2024perturbation} ran this comparison for general task performance. For refusal we know only of a comparison with isotropic noise on first-token margins \citep{razan2026firstorder}, without controls built from the real error.

This comparison is our main experiment, but it needs a reliably measured shift, and individual labels are unstable. An 8-bit build, whose likelihood on reference text is unchanged, produces different text for most prompts and changes the judged label of several percent of them (\S\ref{sec:instrument}). We therefore report every result as a paired shift in a rate, alongside the same shift for the 8-bit build. We test two settings. In the \emph{forced start} (\S\ref{sec:forced}), the first tokens of a compliant reply are prefilled, which tests how easily refusal is bypassed. In the \emph{free start} (\S\ref{sec:free}), the model receives held-out in-the-wild jailbreak prompts and chooses how to begin.

\begin{figure*}[t]
\centering
\includegraphics[width=\textwidth]{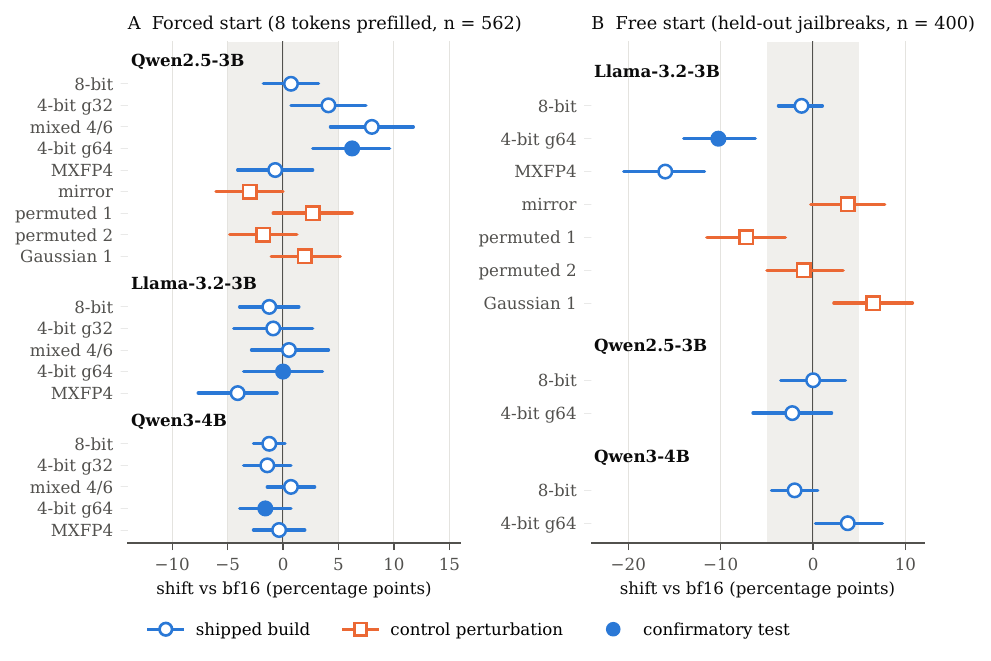}
\caption{Shift in judged harmful output (A) or fulfilment (B) relative to bf16, with 95\% paired bootstrap intervals. The shaded band is the $\pm$5\pp{} equivalence margin. Filled markers are confirmatory tests. Three of six random controls are shown (all six: Figure~\ref{fig:controls}). Blue circles are shipped builds, and orange squares are matched-perturbation controls built from the 4-bit rounding error (\S\ref{sec:controls}). The intervals of all 8-bit builds include zero. \qwen's affine 4-bit builds shift under a forced start, and \llama's 4-bit and MXFP4 builds shift under a free start.}
\label{fig:forest}
\end{figure*}

We find two shifts, with opposite signs (Figure~\ref{fig:forest}). With a forced start, 4-bit \qwen{} continues a harmful reply more often than the original. 4-bit \llama{} does not, but in the free start it complies with fewer jailbreak prompts. This is not better discrimination of harmful requests, because it also refuses more benign requests that are written like jailbreaks (\S\ref{sec:interp}). \qwenthree{} shows no significant shift in either setting.

We then test whether each shift depends on the rounding error or only on its size (\S\ref{sec:controls}). Shuffled or Gaussian errors of the same size move the models by up to several points in either direction, but their mean is close to zero. Flipping the sign of the real error flips the sign of the estimate, and scaling the error changes behaviour near-linearly. The direction of each shift therefore depends on the specific rounding error and is not the model's general response to a perturbation of that size. In the activations (\S\ref{sec:direction}), the real error moves both models along the direction that mediates refusal, but so do builds whose behaviour is unchanged, so this projection does not predict a behavioural shift. Among other quantizers, AWQ reproduces \qwen's shift but not \llama's, and llama.cpp's Q4\_K\_M shifts neither model, although an MLX build with the same bits per weight does shift \qwen.

Our scope is narrow: three small models, greedy decoding, and 4-bit builds from \texttt{mlx-lm} and llama.cpp. Within it, neither bits per weight nor likelihood predicts whether a build's refusal behaviour changes, so each build has to be compared with its original, alongside a null.

\paragraph{Contributions.}
We contribute (i) a measurement of two 4-bit shifts in refusal, with opposite signs; (ii) their attribution to the specific rounding error by matched perturbations and a dose-response; (iii) a refusal-direction analysis; and (iv) a paired design with an 8-bit null for comparing a model with any variant of it. We release code, prompt identifiers and labels.

\section{Related work}
\label{sec:related}

\paragraph{Refusal in quantized models.}
Audits of quantized chat models report effects whose sign differs between models and methods \citep{hong2024decoding, chen2025qresafe, kharinaev2025investigating, kadadekar2026quality, prasad2026temperature, wu2026decisions}. One audit covers 66 quantized variants, scored by an LLM judge checked against human labels, and finds no method consistently safest \citep{kharinaev2025investigating}. In another, AWQ, GPTQ and 2--3-bit GGUF builds changed refusal by up to 57 points, while no Q4\_K\_M build changed it by more than 3 \citep{kadadekar2026quality}. Quantization can also be attacked, so that a model behaves well at full precision and badly once quantized \citep{egashira2024exploiting, egashira2025gguf}. None of these studies perturbs the weights by a matched non-quantization error.

\paragraph{Quantization as perturbation.}
\citet{gong2024perturbation} treat quantization as noise on weights and activations and compare it with artificial perturbations of matched L2 norm. For general performance, they find that the two agree in some model families and differ in others. We ask the same question about refusal, and add controls built from the real error (its mirror image, shuffles and scaled copies), which separate its pattern and sign from its size. For refusal, \citet{razan2026firstorder} compare isotropic weight noise with real 4-bit error on first-token margins: the noise model under-predicts the flips, and the refusals held most weakly at full precision are the ones that break. We use full replies and controls built from the real error. Random weight perturbations leave a model's average safety intact within a basin around it \citep{peng2024safety}.

\paragraph{Instability of single labels.}
Quantized models flip individual answers while accuracy stays flat \citep{dutta2024accuracy}. Safety judgments flip across seeds and temperatures \citep{larsen2025instability, prasad2026temperature}, attack success rates vary from run to run \citep{greatpretender2026}, and noise added to activations \citep{shahani2025noise} or parameters \citep{liu2026aligned} erodes safety alignment, without being matched to a quantizer's error. We treat this instability as a baseline and define our estimand with it in mind (\S\ref{sec:instrument}).

\paragraph{Prefill and shallow alignment.}
Refusal is concentrated in the first few tokens of a reply \citep{qi2024shallow} and is mediated by a single direction in the residual stream \citep{arditi2024refusal}. Prefilling a compliant start bypasses it in full-precision open models \citep{qi2024shallow, struppek2026prefill}. \citet{park2026quantization} write compliant prefixes into the assistant turn of GGUF builds and find that failures at 2 bits concentrate in the smallest models; \citet{chhabra2025towards} find that the refusal direction itself survives 8-bit and AWQ quantization (cosine similarity above 0.99). We use prefill depth as a probe against the full-precision original, and ask whether the rounding error moves activations along that direction.

\section{Setup}
\label{sec:setup}

\paragraph{Scope.}
A developer evaluates a chat model at full precision and ships a 4-bit copy, and we ask whether the copy refuses the same requests. We use assistant prefill, which forces the first tokens of the reply, to test how easily refusal is bypassed \citep{qi2024shallow}. It is not a realistic attack on local weights, whose holder can remove safety training directly.

\paragraph{Models and formats.}
We use \qwen-Instruct, \llama-Instruct and \qwenthree-Instruct-2507 with their default chat templates, pinning Llama's template date so that runs are comparable. Every quantized build is converted from the bf16 checkpoint with \texttt{mlx-lm} 0.32.2. Our primary format is the \texttt{mlx-lm} default, affine round-to-nearest at 4 bits with groups of 64. Around it we build 8 bits, 4 bits with groups of 32, \texttt{mlx-lm}'s mixed 4/6-bit recipe and MXFP4. We also build \texttt{mlx-lm}'s AWQ and GPTQ \citep{lin2024awq, frantar2023gptq}, which write the same 4-bit, group-64 format as the default, so their bits per weight are the same. For a second runtime we build bf16, Q8\_0 and Q4\_K\_M GGUF files with llama.cpp from the same checkpoints and send it the same prompts as token ids. Table~\ref{tab:formats} gives each format's bits per weight and its capability loss, measured as the change in negative log-likelihood (NLL) that the build assigns to bf16's own replies to 200 harmless prompts.

\begin{table}[t]
\centering\small
\setlength{\tabcolsep}{3.5pt}
\begin{tabular}{lcccc}
\toprule
Format & bits/w & \multicolumn{3}{c}{$\Delta$NLL/token vs bf16} \\
 & & Qwen2.5 & Llama & Qwen3 \\
\midrule
8bit-g64   & \N{cap.qwen.8bit-g64.bpw}  & \N{cap.qwen.8bit-g64.dnll}  & \N{cap.llama.8bit-g64.dnll}  & \N{cap.qwenthree.8bit-g64.dnll} \\
4bit-g32   & \N{cap.qwen.4bit-g32.bpw}  & \N{cap.qwen.4bit-g32.dnll}  & \N{cap.llama.4bit-g32.dnll}  & \N{cap.qwenthree.4bit-g32.dnll} \\
mixed\_4\_6 & \N{cap.qwen.mixed46.bpw} & \N{cap.qwen.mixed46.dnll}   & \N{cap.llama.mixed46.dnll}   & \N{cap.qwenthree.mixed46.dnll} \\
4bit-g64   & \N{cap.qwen.4bit-g64.bpw}  & \N{cap.qwen.4bit-g64.dnll}  & \N{cap.llama.4bit-g64.dnll}  & \N{cap.qwenthree.4bit-g64.dnll} \\
MXFP4      & \N{cap.qwen.mxfp4.bpw}     & \N{cap.qwen.mxfp4.dnll}     & \N{cap.llama.mxfp4.dnll}     & \N{cap.qwenthree.mxfp4.dnll} \\
AWQ        & \N{cap.qwen.awq.bpw}       & \N{cap.qwen.awq.dnll}       & \N{cap.llama.awq.dnll}       & -- \\
GPTQ       & \N{cap.qwen.gptq.bpw}      & \N{cap.qwen.gptq.dnll}$^\dagger$ & \N{cap.llama.gptq.dnll}      & -- \\
\bottomrule
\end{tabular}
\caption{Formats. Bits per weight include scales and biases. $\Delta$NLL is the change in mean per-token negative log-likelihood of bf16's own replies to 200 harmless prompts (bf16: 0.327, 0.203, 0.148). The 8-bit builds change likelihood very little, and MXFP4 loses the most in every model. AWQ and GPTQ were built only for the two models with a shift. $^\dagger$The \qwen{} GPTQ build failed and is excluded from all analyses (Appendix~\ref{app:prereg}). The llama.cpp builds (bf16, Q8\_0, Q4\_K\_M at \N{S9.qwenforced.gguf-q80.bpw} and \N{S9.qwenforced.gguf-q4km.bpw} bits per weight) run in a different runtime and are not comparable on this measure. They are reported in \S\ref{sec:controls} against their own bf16.}
\label{tab:formats}
\end{table}

\paragraph{Two settings.}
In the \emph{forced start}, the model sees one of the \N{C1.permodel.qwen.all.n} unique requests of the \texttt{harmful\_test} split of \citet{arditi2024refusal}, and the first $k=8$ tokens of a compliant reply are prefilled as token ids. The prefix comes from an abliterated copy of the same bf16 model \citep{arditi2024refusal}, so every build of a model continues the same tokens. We also report $k \in \{0, 2\}$ and a generic prefix, ``Sure, here is''. In the \emph{free start}, the model sees adversarial-harmful prompts from the WildJailbreak evaluation set \citep{jiang2024wildjailbreak}, which are in-the-wild jailbreak tactics, with no prefill. The first 200 prompts form a development set, and 400 further prompts, disjoint from these, form the held-out test set. Over-refusal is measured on XSTest's 250 safe prompts \citep{rottger2024xstest} and on WildJailbreak's 210 adversarial-benign prompts, which are benign requests written in jailbreak style. SORRY-Bench \citep{xie2024sorry} is a secondary family (Appendix~\ref{app:families}). Decoding is greedy with a cap of 128 new tokens.

\paragraph{Judges and human labels.}
The outcome is whether the reply's content does what was asked. A content judge (Qwen2.5-14B-Instruct) reads the request and the full reply and labels it as complying, partly complying, not complying, or incoherent. The first two count as \emph{judged harmful} in the forced start and as \emph{fulfilment} in the free start, where some prompts are mild. Keyword refusal detection cannot be used here, because the prefix fills the part of the reply that it reads. Llama Guard~3 \citep{llamaguard3} is a second judge with a different construct (``unsafe''). For the forced-start primary test we add human labels in a two-phase design: every pair of replies that the judge labelled differently, plus 25 random pairs from each of the two concordant strata, shown without format information and in random order, and design-weighted. The author labelled \N{humanpairs.total} pairs (\N{human.totalreplies} replies). Appendix~\ref{app:judge} gives the rubric and protocol.

\paragraph{Statistics.}
Every comparison is paired, with the same unique prompts under two builds. The \emph{shift} is the difference in rates, and \emph{churn} is the share of prompts whose label differs. Tests are exact McNemar tests, Holm-corrected within each family. Intervals are percentile bootstrap intervals over prompts with 10{,}000 resamples. Equivalence means that the 90\% interval of the shift lies within $\pm$5\pp{}. The primary and confirmatory tests, the secondary families and the decision rules were registered before the data they concern; Appendix~\ref{app:prereg} lists what was registered when, what changed, and which analyses are exploratory.

\section{Label instability and the 8-bit null}
\label{sec:instrument}

Before testing whether 4-bit quantization changes refusal, we measured how much judged labels change under perturbations that should not matter. Generation itself is deterministic. Regenerating 100 forced-start replies from the same bf16 weights gives token-identical text in \N{e0.qwen.identical}\% of cases for \qwen{} and \N{e0.llama.identical}\% for \llama, so the runtime does not contribute to the differences reported below. An 8-bit build leaves likelihood on reference text nearly unchanged (\N{cap.qwen.8bit-g64.dnll}, \N{cap.llama.8bit-g64.dnll} and \N{cap.qwenthree.8bit-g64.dnll} nats per token), but only \N{c0.qwen.8bit-g64.identical}\%, \N{c0.llama.8bit-g64.identical}\% and \N{c0.qwenthree.8bit-g64.identical}\% of its forced-start replies match bf16's token for token, and the judged label changes for \N{S6.qwen.8bit-g64.churn}\%, \N{S6.llama.8bit-g64.churn}\% and \N{S6.qwenthree.8bit-g64.churn}\% of prompts. The rate does not change: the 8-bit shifts are \N{S6.qwen.8bit-g64.diff}, \N{S6.llama.8bit-g64.diff} and \N{S6.qwenthree.8bit-g64.diff}\pp{}, all with intervals that include zero. We attribute the changed labels to greedy decoding. Under a forced start many replies are close to the boundary between complying and refusing, a small perturbation changes an early token, and the label of the continuation can then change in either direction. At 4 bits churn roughly doubles, and sampling changes more labels again (Appendix~\ref{app:floor}).

A changed label on a given prompt is therefore not evidence that quantization had an effect, since a perturbation with no net effect changes many labels while leaving the rate where it was. We report each 4-bit result as a paired shift in the rate, compare it with the 8-bit shift, and give churn only as context.

\paragraph{Judge and human labels.}
The judge and the human labels agree on \N{human.qwen.agreedw}\%, \N{human.llama.agreedw}\% and \N{human.qwenthree.agreedw}\% of replies, design-weighted ($\kappa = \N{human.qwen.kappa}$, \N{human.llama.kappa} and \N{human.qwenthree.kappa}). Two of these fall below the registered threshold of 0.60, so in the forced start we lead with the human labels. Their disagreements are mostly in one direction. In \qwen, when the judge labels a reply harmful the human label usually agrees (precision \N{human.qwen.bf16.ppv}--\N{human.qwen.4bit-g64.ppv}\%), but the judge misses many replies that the human labels count as harmful. Design-weighted, it finds harm in \N{human.qwen.judgerate}\% of \qwen's forced-start replies, and the human labels in \N{human.qwen.humanrate}\%. Much of the gap is related to truncation. Generation is capped at 128 tokens, and \N{trunc.qwen.share}\% of \qwen's 4-bit replies reach the cap. Many of these open with a refusal-like preamble, then begin to comply, and are cut off at about that point. The two sources disagree on \N{trunc.dis.truncated}\% of truncated replies and on \N{trunc.dis.complete}\% of replies that end before the cap. The judge tends to require that the harmful content be present in the text, while the human labels count a reply that has clearly begun to deliver it. The disagreement is mostly about replies that are cut off, which is the likely reason the human estimate of the shift is larger.

We find no sign that truncation produces the shift. Among pairs in which both replies end before the cap, the judge-label shift is \N{trunc.qwen.completediff}\pp{} \N{trunc.qwen.completeci} for \qwen{} (\N{C1.permodel.qwen.all.diff}\pp{} overall) and \N{trunc.llama.completediff}\pp{} for \llama{} (\N{C1.permodel.llama.all.diff}\pp{} overall), although only \N{trunc.qwen.ncomplete} and \N{trunc.llama.ncomplete} pairs remain and the \qwen{} interval is wide. The judge, Llama Guard and the human labels agree on which model shifts: \qwen{} shifts toward harmful output, and the other two models show no significant shift (Appendix Table~\ref{tab:labels}). Where there are no human labels we use the judge, which under-counted harm where we could check.

\section{Two 4-bit shifts}
\label{sec:shifts}

\subsection{Forced start: the primary test}
\label{sec:forced}

The forced start tests whether, once a reply has begun to comply, the 4-bit build continues it more often than the original does. It is our primary test because refusal is known to be weakest under prefill \citep{qi2024shallow}, so an erosion of refusal by quantization should be easiest to detect here.

We find such a shift in \qwen{} only. With eight tokens forced, its default 4-bit build continues into harmful content more often than bf16. Under the human labels (\S\ref{sec:instrument}), among the pairs that the judge had labelled differently, the harmful reply came from the 4-bit build in \N{human.qwen.dtofour} and from bf16 in \N{human.qwen.dtobf} ($p=\N{human.qwen.dp}$), and the design-weighted shift is \N{human.qwen.net}\pp{} \N{human.qwen.netci}. The judge, which under-counts harm, gives a smaller shift of the same sign, from \N{C1.permodel.qwen.all.ratea}\% to \N{C1.permodel.qwen.all.rateb}\%. This is a shift of \N{C1.permodel.qwen.all.diff}\pp{} \N{C1.permodel.qwen.all.ci}, with discordant prompts splitting \N{C1.permodel.qwen.all.aonly} to \N{C1.permodel.qwen.all.bonly} (Holm $p=\N{C1.permodel.qwen.all.pholm}$). Llama Guard also gives a positive shift (\N{labels.qwen.llamaguard.diff}\pp{} \N{labels.qwen.llamaguard.ci}).

\llama{} and \qwenthree{} show no shift. Under the judge, both are equivalent to bf16 within the $\pm$5\pp{} equivalence margin (90\% intervals \N{C1.permodel.llama.all.cininety} and \N{C1.permodel.qwenthree.all.cininety}). The human-label estimates, \N{human.llama.net} and \N{human.qwenthree.net}\pp{}, have 90\% intervals too wide to establish equivalence (\N{human.llama.netcininety} and \N{human.qwenthree.netcininety}). All three results hold on the \N{C1.permodel.qwen.nonpilot.n} prompts that were not used in the prefill pilot (a 150-prompt run before registration). Pooled across models the shift is not significant (discordant pairs \N{C1.pooled.aonly} against \N{C1.pooled.bonly}, $p=\N{C1.pooled.p}$), so the effect is specific to \qwen.

The effect also depends on the format. All three affine 4-bit builds of \qwen{} shift in the same direction, by \N{S6.qwen.4bit-g32.diff}\pp{} at group size 32, \N{S6.qwen.4bit-g64.diff}\pp{} at 64, and \N{S6.qwen.mixed46.diff}\pp{} under the mixed 4/6 recipe, which shares most of its tensors with the group-64 build. MXFP4, which loses the most likelihood, shows no shift (\N{S6.qwen.mxfp4.diff}\pp{}), and the four 4-bit builds differ significantly from one another (Cochran's $Q$, Holm $p=\N{S6.qwen.qp}$). The format therefore matters within one model and one bit width, a first indication that the size of a perturbation does not determine its effect. In the other two models no build shifts significantly, apart from a nominal \N{S6.llama.mxfp4.diff}\pp{} for \llama's MXFP4 build.

\qwen's shift is present with two forced tokens (\N{S2.cells.qwen.4bit-g64|k2.diff}\pp{}, Holm $p=\N{S2.holm.qwen.4bit-g64.k2}$), and is smaller with none, where bf16 almost never complies. It persists when the reply cap is tripled (\N{robustness.E4qwen.128.diff}\pp{} at 128 tokens and \N{robustness.E4qwen.384.diff}\pp{} at 384, on a fixed subset of 150 prompts). With the generic prefix ``Sure, here is'' the shift has the same sign but is smaller, \N{robustness.S8.qwen.diff}\pp{} \N{robustness.S8.qwen.ci}, and is nominally significant ($p=\N{robustness.S8.qwen.p}$) but not after Holm correction (\N{robustness.S8.qwen.pholm}).

\subsection{Free start: held-out jailbreak prompts}
\label{sec:free}

In most real misuse there is no forced start. A user sends a jailbreak prompt and the model chooses how to begin its reply. For this setting we use a development set of 200 WildJailbreak prompts and a held-out test set of 400. On the development set (Appendix~\ref{app:pilot}), \llama's 4-bit build was much less compliant than bf16. We test this on the held-out prompts, with every build, an 8-bit null and one confirmatory test with a predicted sign.

The effect holds on the test set. \llama{} fulfils \N{C2.cells.llama.4bit-g64.ratea}\% of the test prompts at bf16 and \N{C2.cells.llama.4bit-g64.rateb}\% at 4 bits, a drop of \N{C2.cells.llama.4bit-g64.absdiff} points \N{C2.cells.llama.4bit-g64.ci}, with discordant prompts splitting \N{C2.cells.llama.4bit-g64.aonly} to \N{C2.cells.llama.4bit-g64.bonly} ($p=\N{C2.cells.llama.4bit-g64.p}$). Its 8-bit build does not differ from bf16 (\N{C2.cells.llama.8bit-g64.diff}\pp{} \N{C2.cells.llama.8bit-g64.ci}), and its MXFP4 build shows a larger drop (\N{C2.cells.llama.mxfp4.diff}\pp{}). The 4-bit drop is the same size when the reply cap is tripled (\N{robustness.E4llama.128.diff}\pp{} at 128 tokens, \N{robustness.E4llama.384.diff}\pp{} at 384), and it has the same sign on SORRY-Bench, a different prompt set with no prefill (\N{S3.llama.4bit-g64.overall.diff}\pp{} \N{S3.llama.4bit-g64.overall.ci}). Llama Guard, which asks a different question (is the reply unsafe?), also finds a drop (\N{labels.freellama.llamaguard.diff}\pp{} \N{labels.freellama.llamaguard.ci}, $p=\N{labels.freellama.llamaguard.p}$).

On the development set \qwenthree's 4-bit build showed an effect of similar size and opposite sign. On the test set it is smaller, \N{C2.cells.qwenthree.4bit-g64.diff}\pp{} \N{C2.cells.qwenthree.4bit-g64.ci}, and not significant after correction. \qwen's shift is \N{C2.cells.qwen.4bit-g64.diff}\pp{} \N{C2.cells.qwen.4bit-g64.ci}.

Table~\ref{tab:settings} compares the two settings. \qwen{} shifts when the start of its reply is forced and not when it is free, and \llama{} shows the reverse. This could reflect whether the prefill or the model opens the reply, but the two settings also use different prompts. To separate the two factors, we also ran the test prompts with a forced generic start (``Sure, here is''). On these prompts \llama's shift is absent (\N{robustness.B1.llama.diff}\pp{} \N{robustness.B1.llama.ci}, $p=\N{robustness.B1.llama.p}$), so on these prompts it depends on \llama{} opening the reply itself. \qwen{} shows no shift on these prompts either (\N{robustness.B1.qwen.diff}\pp{} \N{robustness.B1.qwen.ci}). Its forced-start shift is largest under the abliterated prefix and weaker under ``Sure, here is'' (\S\ref{sec:forced}), so it appears to need a strongly compliant start, and it may also depend on the prompts.

\begin{table}[t]
\centering\small
\begin{tabular}{lcc}
\toprule
4-bit shift (pp) & Forced start & Free start (held out) \\
\midrule
\qwen      & \textbf{\N{C1.permodel.qwen.all.diff}}  & \N{C2.cells.qwen.4bit-g64.diff} \\
\llama     & \N{C1.permodel.llama.all.diff}           & \textbf{\N{C2.cells.llama.4bit-g64.diff}} \\
\qwenthree & \N{C1.permodel.qwenthree.all.diff}       & \N{C2.cells.qwenthree.4bit-g64.diff} \\
\bottomrule
\end{tabular}
\caption{The 4-bit shift in each setting; bold marks a significant confirmatory test. The settings also differ in prompts (\texttt{harmful\_test} with an 8-token prefix vs.\ held-out WildJailbreak with none); a forced generic start on the held-out prompts gives no shift in either model (\S\ref{sec:free}).}
\label{tab:settings}
\end{table}

\section{Two simpler explanations}
\label{sec:interp}

Before testing where the two shifts come from, we consider two simpler explanations. The first is that a quantized model is a generally weaker model, and that refusal degrades along with everything else. Our capability measure, the likelihood that each build assigns to bf16's own harmless replies, does not support this. MXFP4 loses more likelihood than any other format in every model (Table~\ref{tab:formats}), but it does not shift \qwen's forced-start behaviour (\N{S6.qwen.mxfp4.diff}\pp{}). \llama{} loses the least likelihood of the three models at 4 bits (\N{cap.llama.4bit-g64.dnll} nats per token against \N{cap.qwen.4bit-g64.dnll} for \qwen) and has the largest free-start shift. Across formats, the shifts are ordered neither by bits per weight nor by likelihood loss. We did not run a task benchmark, so we cannot say whether one would order them.

The second explanation is that \llama's 4-bit build has become safer, since it complies with fewer jailbreak prompts. It has not (Table~\ref{tab:benign}). On XSTest's plain safe requests, 4-bit \llama{} refuses about as often as bf16 (\N{C4benign.llama.4bit-g64.xstestsafe.diff}\pp{} \N{C4benign.llama.4bit-g64.xstestsafe.ci}). On benign requests \emph{written in the style of a jailbreak}, it refuses more often (\N{C4benign.llama.4bit-g64.wjbbenign.diff}\pp{} \N{C4benign.llama.4bit-g64.wjbbenign.ci}). So the drop is not better discrimination of harm, and some legitimate requests phrased like jailbreaks are now refused (an exploratory comparison, $p=\N{C4benign.llama.4bit-g64.wjbbenign.p}$ uncorrected). The two Qwen models change in the other direction on benign prompts. 4-bit \qwen{} refuses fewer safe XSTest requests (\N{C4benign.qwen.4bit-g64.xstestsafe.diff}\pp{}, Holm $p=\N{C4benign.qwen.4bit-g64.xstestsafe.pholm}$), which is consistent with its shift toward compliance under a forced start, and 4-bit \qwenthree{} refuses fewer jailbreak-styled benign requests (\N{C4benign.qwenthree.4bit-g64.wjbbenign.diff}\pp{}).

\begin{table}[t]
\centering\small
\setlength{\tabcolsep}{4pt}
\begin{tabular}{llcc}
\toprule
 & & XSTest safe & WJB benign \\
\midrule
\multirow{2}{*}{\qwen} & bf16 & \N{C4benign.qwen.4bit-g64.xstestsafe.ratea}\% & \N{C4benign.qwen.4bit-g64.wjbbenign.ratea}\% \\
 & 4-bit $\Delta$ & \N{C4benign.qwen.4bit-g64.xstestsafe.diff} & \N{C4benign.qwen.4bit-g64.wjbbenign.diff} \\
\multirow{2}{*}{\llama} & bf16 & \N{C4benign.llama.4bit-g64.xstestsafe.ratea}\% & \N{C4benign.llama.4bit-g64.wjbbenign.ratea}\% \\
 & 4-bit $\Delta$ & \N{C4benign.llama.4bit-g64.xstestsafe.diff} & \N{C4benign.llama.4bit-g64.wjbbenign.diff} \\
\multirow{2}{*}{\qwenthree} & bf16 & \N{C4benign.qwenthree.4bit-g64.xstestsafe.ratea}\% & \N{C4benign.qwenthree.4bit-g64.wjbbenign.ratea}\% \\
 & 4-bit $\Delta$ & \N{C4benign.qwenthree.4bit-g64.xstestsafe.diff} & \N{C4benign.qwenthree.4bit-g64.wjbbenign.diff} \\
\bottomrule
\end{tabular}
\caption{Refusal (partial or full) of benign prompts: the bf16 rate and the 4-bit shift in pp, where a positive shift is more refusal. WJB benign prompts are benign requests written in jailbreak style. MXFP4 rows in Appendix~\ref{app:families}.}
\label{tab:benign}
\end{table}

\section{Matched-perturbation controls}
\label{sec:controls}

Neither shift is explained by lost capability (\S\ref{sec:interp}). This section tests what determines the direction of each. If other perturbations of the same size move a model in the same direction as its 4-bit build does, the direction is a property of the model and would appear under any such perturbation. If they do not, the direction depends on the pattern of the rounding error.

\paragraph{Matched perturbations.}
For each weight matrix $W$ that the 4-bit build quantizes, embedding included, let $E = \mathrm{dequant}(Q(W)) - W$ be its rounding error. We then build fake rounding errors: dense bf16 models that perturb $W$ by the same amount as rounding does, in a different pattern. The \emph{mirror}, $W - E$, reverses the sign of the error. Three \emph{permuted} builds shuffle $E$ within each quantization group of 64, so every group keeps the same set of error values. Three \emph{Gaussian} builds replace $E$ with noise scaled to each row's norm. In every control each row is perturbed by the same norm as under rounding, and the controls lose a similar or somewhat larger amount of likelihood (Appendix~\ref{app:controls}). A \emph{dense} build, $W + E$, checks the construction. Its likelihood differs from the real quantized model's by \N{c3.qwen.gate.nlldiff} and \N{c3.llama.gate.nlldiff} nats per token, and its first token matches on \N{c3.qwen.gate.agree}\% and \N{c3.llama.gate.agree}\% of prompts. It also reproduces both shifts: \N{C3.qwenforced.dense.diff}\pp{} \N{C3.qwenforced.dense.ci} for \qwen{} and \N{C3.llamafree.dense.diff}\pp{} \N{C3.llamafree.dense.ci} for \llama, within \N{C3.qwenforced.densevsreal.absdiff} and \N{C3.llamafree.densevsreal.absdiff} points of the real builds.

We classify the direction of each shift with a decision rule. The \emph{random-control shift} is the mean shift over the permuted and Gaussian builds. A direction is the \emph{model's lean} if the random controls share the real build's sign with an interval excluding zero and the mirror does not reverse it. It is \emph{rounding-specific} if the mirror reverses it with an interval excluding zero, or if the random-control interval includes zero while the real-minus-random interval does not.

\IfMacro{C3.llamafree.summary.verdict}{Under this rule, applied to all six random draws, both shifts are \N{C3.qwenforced.summary.verdict}. For \qwen, the six random controls average \N{C3.qwenforced.summary.random}\pp{} \N{C3.qwenforced.summary.random.ci}, the real build exceeds them by \N{C3.qwenforced.summary.realminusrandom}\pp{} \N{C3.qwenforced.summary.realminusrandom.ci}, and the mirror's estimate has the opposite sign (\N{C3.qwenforced.e2-mirror.diff}\pp{} \N{C3.qwenforced.e2-mirror.ci}). \llama{} shows the same pattern with the signs reversed: random controls \N{C3.llamafree.summary.random}\pp{} \N{C3.llamafree.summary.random.ci}, real minus random \N{C3.llamafree.summary.realminusrandom}\pp{} \N{C3.llamafree.summary.realminusrandom.ci}, mirror \N{C3.llamafree.e2-mirror.diff}\pp{} \N{C3.llamafree.e2-mirror.ci}. Neither mirror is significant by itself, but each differs clearly from its real build, and half of that difference is \N{C3.qwenforced.summary.odd}\pp{} \N{C3.qwenforced.summary.odd.ci} and \N{C3.llamafree.summary.odd}\pp{} \N{C3.llamafree.summary.odd.ci}. In \llama{} the two random families differ: the permuted builds shift by \N{C3.llamafree.e2-perm1.diff}, \N{C3.llamafree.e2-perm2.diff} and \N{C3.llamafree.e2-perm3.diff}\pp{} and the Gaussian builds by \N{C3.llamafree.e2-gauss.diff}, \N{C3.llamafree.e2-gauss2.diff} and \N{C3.llamafree.e2-gauss3.diff}\pp, and the part of \llama's shift that does not change sign with the error, \N{C3.llamafree.summary.even}\pp{} \N{C3.llamafree.summary.even.ci}, is up to a third of it.}{Both shifts are rounding-specific. For \qwen, the random controls average \N{C3.qwenforced.summary3.random}\pp{} \N{C3.qwenforced.summary3.random.ci}, the real build exceeds them by \N{C3.qwenforced.summary3.realminusrandom}\pp{} \N{C3.qwenforced.summary3.realminusrandom.ci}, and the mirror's estimate has the opposite sign (\N{C3.qwenforced.e2-mirror.diff}\pp{} \N{C3.qwenforced.e2-mirror.ci}). \llama{} shows the same pattern with the signs reversed: random controls \N{C3.llamafree.summary3.random}\pp{} \N{C3.llamafree.summary3.random.ci}, real minus random \N{C3.llamafree.summary3.realminusrandom}\pp{} \N{C3.llamafree.summary3.realminusrandom.ci}, mirror \N{C3.llamafree.e2-mirror.diff}\pp{} \N{C3.llamafree.e2-mirror.ci}. Neither mirror is significant by itself, but each differs clearly from its real build, and half of that difference is \N{C3.qwenforced.summary.odd}\pp{} \N{C3.qwenforced.summary.odd.ci} and \N{C3.llamafree.summary.odd}\pp{} \N{C3.llamafree.summary.odd.ci}. In \llama{} the two random families differ: the permuted builds shift by \N{C3.llamafree.e2-perm1.diff}, \N{C3.llamafree.e2-perm2.diff} and \N{C3.llamafree.e2-perm3.diff}\pp{} and the Gaussian builds by \N{C3.llamafree.e2-gauss.diff}, \N{C3.llamafree.e2-gauss2.diff} and \N{C3.llamafree.e2-gauss3.diff}\pp, and the part of \llama's shift that does not change sign with the error, \N{C3.llamafree.summary.even}\pp{} \N{C3.llamafree.summary.even.ci}, is up to a third of it.}

\begin{figure}[t]
\centering
\includegraphics[width=\columnwidth]{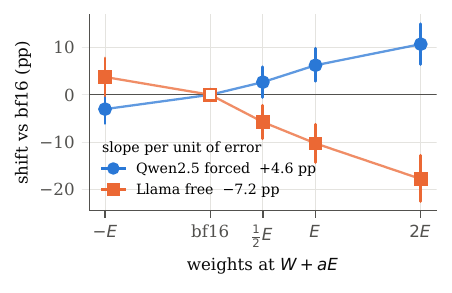}
\caption{Dose. Each build is $W + aE$, where $E$ is the real 4-bit rounding error; $a=-1$ is the mirror and $a=1$ the real build. Shift against bf16 with 95\% intervals; the legend gives the slope. Every control build is in Appendix~\ref{app:controls}.}
\label{fig:dose}
\end{figure}

\paragraph{Dose.}
The mirror is the real error scaled by $-1$. Builds at $\tfrac12$ and $2$ extend it to a line through weight space, $W + aE$ (Figure~\ref{fig:dose}). In both models the shift is monotone in $a$ and near-linear over the tested range, and its estimate has the opposite sign at $a=-1$. The slopes are \N{C3.qwenforced.slope}\pp{} \N{C3.qwenforced.slope.ci} and \N{C3.llamafree.slope}\pp{} \N{C3.llamafree.slope.ci} per unit of error. The doubled builds lose much more likelihood than the real builds (\N{C3.qwenforced.e2-scale2.nllgap} and \N{C3.llamafree.e2-scale2.nllgap} nats per token, against \N{C3.qwenforced.realnllgap} and \N{C3.llamafree.realnllgap}), and the conclusion does not depend on them. The half-size build of \llama{} is more informative. Its change in likelihood (\N{C3.llamafree.e2-scale0.5.nllgap} nats per token) is probably too small for a perplexity check to detect, and it is already \N{C3.llamafree.e2-scale0.5.absdiff} points less compliant ($p=\N{C3.llamafree.e2-scale0.5.p}$).

\paragraph{Other quantizers and a second runtime.}
Other quantizers produce real rounding errors with a different structure at the same bit width. AWQ and GPTQ write the default format at \N{cap.qwen.awq.bpw} bits per weight but choose the quantized values using calibration data, and llama.cpp's Q4\_K\_M uses a different scheme in a different runtime. AWQ reproduces \qwen's shift (\N{S9.qwenforced.awq.diff}\pp{} \N{S9.qwenforced.awq.ci}, against the default build's \N{S9.qwenforced.4bit-g64.diff}) but not \llama's (\N{S9.llamafree.awq.diff}\pp{} \N{S9.llamafree.awq.ci}). Q4\_K\_M shifts neither model, and in both cells its estimate has the \emph{opposite} sign to the default build's: \N{S9.qwenforced.gguf-q4km.diff}\pp{} \N{S9.qwenforced.gguf-q4km.ci} and \N{S9.llamafree.gguf-q4km.diff}\pp{} \N{S9.llamafree.gguf-q4km.ci}. For \qwen{} its larger bit budget does not explain this, because at \N{S9.qwenforced.gguf-q4km.bpw} bits per weight it matches our \texttt{4bit-g32}, which does shift \qwen{} (\N{S6.qwen.4bit-g32.diff}\pp{}). Q4\_K\_M still changes \N{S9.qwenforced.gguf-q4km.churn}\% and \N{S9.llamafree.gguf-q4km.churn}\% of labels, without a net direction. The two runtimes do not differ significantly at bf16 (\N{S9.qwenforced.runtimebf16.diff}\pp, $p=\N{S9.qwenforced.runtimebf16.p}$; \N{S9.llamafree.runtimebf16.diff}\pp, $p=\N{S9.llamafree.runtimebf16.p}$). GPTQ contributes one cell, \llama{} at \N{S9.llamafree.gptq.diff}\pp{} \N{S9.llamafree.gptq.ci} (Holm $p=\N{S9.llamafree.gptq.pholm}$), with the same sign as the default build. The \qwen{} GPTQ build failed and is excluded, and \llama's Q8\_0 build is nominally lower but does not survive Holm correction (both in Appendix~\ref{app:prereg}). Builds at the same bit width thus differ in whether they shift a model.

\section{Refusal-direction projections}
\label{sec:direction}

The controls show that each shift depends on its rounding error, but not what that error changes inside the model. Refusal in these models is mediated by a single direction in the residual stream \citep{arditi2024refusal}, the same direction that we use to build the compliant prefixes. For every build, on the prompts of the behavioural tests, we record the projection of the model's state at the point of generation onto its bf16 refusal direction, and compare it with bf16 prompt by prompt. The analysis is correlational and identifies no circuit. We test two predictions. First, 4-bit \qwen{} under a forced start should have a \emph{lower} projection than bf16, and 4-bit \llama{} under a free start a \emph{higher} one. Second, across all builds of a cell, real and control, the change in projection should be negatively correlated with the change in compliance.

\qwen{} fits the first prediction but not the second. At the layer used for abliteration (layer \N{M1.qwenforced.layer}), the projection of its 4-bit build under a forced start is lower than bf16's (\N{M1.qwenforced.4bit-g64.mean} \N{M1.qwenforced.4bit-g64.ci}), and the mirror build's is higher (\N{M1.qwenforced.e2-mirror.mean}). The projection scales with the dose, at \N{M1.qwenforced.e2-scale0.5.mean}, \N{M1.qwenforced.4bit-g64.mean} and \N{M1.qwenforced.e2-scale2.mean} for $a = \tfrac12, 1, 2$. Across the \N{M1.qwenforced.nbuilds} builds of this cell, however, the two are not significantly correlated (Spearman $\rho=\N{M1.qwenforced.rho}$, $p=\N{M1.qwenforced.rhop}$). Several builds lower the projection nearly as much as the default build without making \qwen{} more compliant, among them a Gaussian control (\N{M1.qwenforced.e2-gauss2.mean}) and MXFP4 (\N{M1.qwenforced.mxfp4.mean}). The same 4-bit build lowers the projection further under a free start (\N{M1.qwenfree.4bit-g64.mean} \N{M1.qwenfree.4bit-g64.ci}), where its behaviour does not shift.

\llama{} fits both predictions, but weakly. The projection of its 4-bit build is slightly higher than bf16's, as predicted (\N{M1.llamafree.4bit-g64.mean} \N{M1.llamafree.4bit-g64.ci}), and the mirror's is lower (\N{M1.llamafree.e2-mirror.mean}). Across \N{M1.llamafree.nbuilds} builds the two are correlated as predicted ($\rho=\N{M1.llamafree.rho}$, $p=\N{M1.llamafree.rhop}$). However, the doubled build, which has the largest behavioural shift, changes the projection in the opposite direction (\N{M1.llamafree.e2-scale2.mean}), and under a forced start, where \llama's behaviour does not shift, the 4-bit build changes the projection by about as much (\N{M1.llamaforced.4bit-g64.mean} \N{M1.llamaforced.4bit-g64.ci}).

In both models the real error moves the state at the start of the reply along the refusal direction with the predicted sign, and the mirror moves it the other way, but this also happens in builds and settings whose behaviour does not change.

\section{Conclusion}
Quantizing these chat models to 4 bits did not change refusal in one direction. \qwen{} became more likely to continue a compliant reply that was started for it, \llama{} became less compliant with jailbreak prompts and with benign prompts in the same style, and \qwenthree{} showed no significant shift. Matched random perturbations moved the models in both directions and averaged close to zero, while each shift had the sign of the real rounding error and grew with its scale over the tested range. Because each build carries its own rounding error, audits of different builds can disagree even about the same model.

\paragraph{What a practitioner can do.}
Where memory allows, use 8 bits: no 8-bit build shifted a rate significantly after correction. At 4 bits, a finer group size and the mixed 4/6 recipe did not remove \qwen's shift, and the half-size \llama{} build shows that a likelihood check can miss one. Q4\_K\_M shifted neither model, but nothing we measured predicts which builds shift, so test each one against its original.

\clearpage
\section*{Limitations}
\begin{itemize}
\setlength\itemsep{1pt}
\item \textbf{Scope.} We study three 3--4B models from two vendors. Because the signs differ between models, the results do not generalize to 4-bit quantization of chat models as a class. Each result applies to the build that was tested.
\item \textbf{Quantizers and runtimes.} We cover MLX round-to-nearest (affine and MXFP4), MLX's AWQ and GPTQ, and llama.cpp's Q4\_K\_M without an importance matrix. Other implementations, imatrix GGUF builds, bitsandbytes and vendor pipelines (quantization-aware training) are not covered. The second runtime and the calibrated quantizers were run only in the two cells with a shift.
\item \textbf{The free-start effect was found on the development set.} The test prompts are held out, but the choice of prompt set and model for the confirmatory test followed the development-set result.
\item \textbf{Two different prompt sets.} The forced-start and free-start settings use different prompts. A forced generic start on the free-start prompts separates the two for \llama{} but not for \qwen, whose shift may depend on prompt type as well as on the start. That forced start also raises baseline compliance, so the comparison is not exactly like for like.
\item \textbf{The dose builds are not real formats.} Builds at half and twice the rounding error are dense bf16 models that no shipped quantizer produces, and the doubled builds lose much more likelihood than the real ones.
\item \textbf{The controls are coarse.} We built one mirror, three permutation seeds, three Gaussian draws and two scales per cell, in two cells. Six random draws give a rough estimate of the distribution of a matched perturbation's effect, the decision rule pools permuted and Gaussian draws although in \llama{} the permuted draws alone average about half the real shift, and the controls attribute direction to the whole perturbation, not to particular layers. The refusal-direction analysis is correlational and identifies no circuit.
\item \textbf{Decoding.} Decoding is greedy with one reply per prompt, except in the seed-floor runs, whereas deployed models sample. Replies are capped at 128 tokens and many reach the cap. We checked the cap on 150 prompts per cell.
\item \textbf{The forced-start prefix is model-specific.} It comes from an abliterated full-precision model and is read by the judge. A generic ``Sure, here is'' prefix gives the same sign for \qwen{} and no shift in the other two.
\item \textbf{Judges and labels.}
\begin{itemize}
\setlength\itemsep{0pt}
\item \textbf{Truncated replies.} Most replies reach the 128-token cap, and the two label sources treat a reply cut off mid-compliance differently (\S\ref{sec:instrument}). We report both. The sign of the effect does not depend on the choice, but its size does: the judge gives \N{C1.permodel.qwen.all.diff}\pp{} and the human labels \N{human.qwen.net}\pp{} for the same contrast, and the human interval is wide.
\item The main judge under-counts harm relative to the human labels. Only the forced-start primary test has human labels, and every other cell uses the judge, which under-counted harm where we could check.
\item The main judge is itself a 4-bit model, of the same family as one subject. The second judge is also quantized and measures a different construct.
\item The only human annotator is the author, who labelled the two replies of each pair side by side.
\end{itemize}
\item \textbf{Seed floor.} It used the same sampler for both models, not each model's shipped settings, and was not run for \qwenthree.
\item \textbf{Capability.} Our capability measure is likelihood on bf16's own replies, which favours bf16. We ran no task benchmark.
\end{itemize}

\section*{Ethical considerations}
This work measures how often quantized copies of public chat models comply with harmful requests. It introduces no new attack: prefill is a known way to bypass refusal, and quantization is routine. We release code and every judge and human label keyed by prompt index, and we withhold the models' replies to harmful prompts. The human labels were produced by the author, who read harmful replies and could stop at any time.

\bibliography{refs}
\appendix
\section{Pre-registration}
\label{app:prereg}
Each registered test was fixed before the data it concerns (Table~\ref{tab:prereg}). The plan is \texttt{PREREG.md} in the repository; it was kept as dated commits in a private development repository, not deposited with a public registry, and that history is available on request.

\begin{table}[h]
\centering\small
\begin{tabular}{@{}lp{0.74\columnwidth}@{}}
\toprule
Date (2026) & Registered \\
\midrule
Nov 10 & forced-start primary, depth, judge, churn, format ladder, seed floor \\
Nov 11 & pooled test, two-phase human labels \\
Nov 13 & free-start test on held-out prompts; matched-perturbation controls \\
Nov 14 & other quantizers, dose, refusal direction, GGUF; six random draws; same-prompt bridge \\
\bottomrule
\end{tabular}
\caption{What was registered when.}
\label{tab:prereg}
\end{table}

\paragraph{Changes to the plan.}
\begin{itemize}[leftmargin=*,itemsep=1pt,topsep=2pt]
\item \emph{Capability} is the NLL each build assigns to bf16's replies to harmless prompts (Table~\ref{tab:formats}), not MMLU.
\item \emph{Seed floor:} one sampler for both models (T 0.7, top-$p$ 0.8, top-$k$ 20); not run for \qwenthree.
\item \emph{Length check} on a fixed random subset of 150 prompts.
\item \emph{SORRY-Bench} intervals are clustered over its 120 questions.
\item \emph{Dense $W+E$} was registered, and first run, on the first 150 prompts of each cell. There the \qwen{} dense build shifted by $+0.7$\pp{} against $+4.7$ for the real build (difference $-4.0$, $[-9.3, +1.3]$), which 150 prompts could not resolve, so we reran it on every prompt (\S\ref{sec:controls}).
\item \emph{Human labels:} the judge-failure criterion ($\kappa<0.60$) was met in two models. The human-label test is an exact test on the judge-discordant stratum with a stratified bootstrap interval for the weighted shift, not the registered McNemar test on weighted counts, and a further 100 concordant pairs planned for this case were not labelled.
\item \emph{Exploratory, uncorrected:} refusal of WildJailbreak's benign prompts (run with the development set), the complete-replies subset (\S\ref{sec:instrument}) and Llama Guard on the forced start were not registered.
\item \emph{Registered and not reported here:} intervals for the risk ratios and the pooled shift, the depth interaction, the prefix-only baseline, the prefix-exclusion and turn-marker sensitivity outcomes, first-32-token identity and per-layer projection profiles. \qwenthree's forced cell was run through the mirror, two permuted and one Gaussian control as registered; all four are within 1.1\pp{} of bf16. The verdict on the three random draws registered first is the same (\N{C3.qwenforced.summary3.verdict}; \N{C3.llamafree.summary3.verdict}).
\end{itemize}

\paragraph{GPTQ.} \qwen's GPTQ build loses \N{cap.qwen.gptq.dnll} nats per token against bf16 (AWQ: \N{cap.qwen.awq.dnll}) and is excluded from every analysis. A build with 6-bit fallback layers (\N{cap.qwen.gptq6.bpw} bits per weight) is equally broken (\N{cap.qwen.gptq6.dnll}), so the failure lies in the GPTQ pass on this model. \llama's GPTQ build is sound (\N{cap.llama.gptq.dnll}) and is reported.

\paragraph{\llama's Q8\_0 build.} It is \N{S9.llamafree.gguf-q80.absdiff} points less compliant than its own bf16 build on the free start (\N{S9.llamafree.gguf-q80.ci}, $p=\N{S9.llamafree.gguf-q80.p}$), the only nominally significant result among the four GGUF comparisons; it does not survive Holm (\N{S9.llamafree.gguf-q80.pholm}), and the MLX 8-bit build shows no shift on the same prompts.

\section{Text identity, churn and the sampling floor}
\label{app:floor}
\begin{figure*}[h]
\centering
\includegraphics[width=\textwidth]{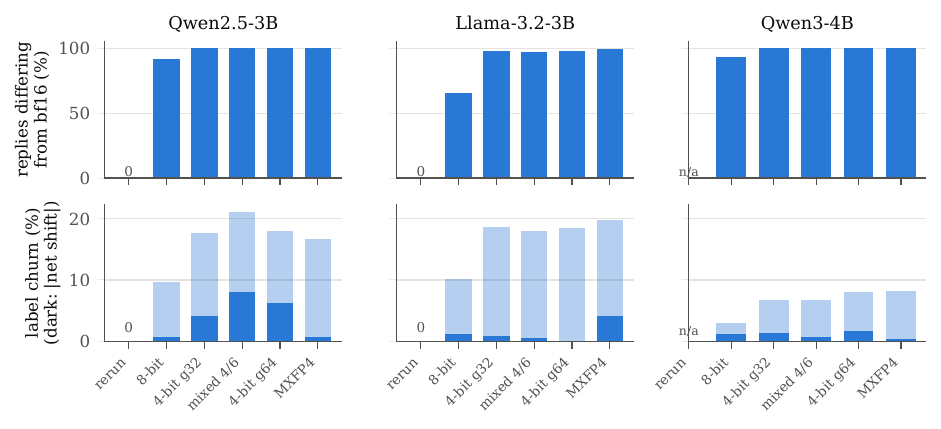}
\caption{The ladder of perturbations, forced-start setting ($k=8$, $n=562$). Top: share of replies whose text differs from bf16's. Bottom: share of prompts whose judged label differs (light) and the absolute net shift in the rate (dark). A greedy rerun of the same weights changes nothing; an 8-bit build, with unchanged likelihood, rewrites most replies and flips 3--10\% of labels while leaving the rate in place.}
\label{fig:ladder}
\end{figure*}

At 4 bits, churn roughly doubles relative to 8 bits (\N{S6.qwen.4bit-g64.churn}\%, \N{S6.llama.4bit-g64.churn}\% and \N{S6.qwenthree.4bit-g64.churn}\% of forced-start prompts). Under sampling, two runs of the same bf16 weights disagree on \N{S7.qwen.floor}\% (\qwen) and \N{S7.llama.floor}\% (\llama) of prompts, and a bf16 run and a 4-bit run disagree on \N{S7.qwen.quant}\% and \N{S7.llama.quant}\%. The excess is \N{S7.qwen.diff}\pp{} \N{S7.qwen.ci} and \N{S7.llama.diff}\pp{} \N{S7.llama.ci}.

Three sampled runs of each build give the per-prompt estimand directly: the per-prompt variance of the label, averaged over the \N{e3.qwen.n} prompts. For \llama{} the 4-bit build's variance (\N{e3.llama.varq}) barely exceeds bf16's (\N{e3.llama.varbf}); the excess is \N{e3.llama.excess} \N{e3.llama.ci}, so quantization does not make its individual labels less stable; with the churn comparison above, its 4-bit build behaves under a forced start like another sample of the original. For \qwen{} the excess is \N{e3.qwen.excess} \N{e3.qwen.ci} (\N{e3.qwen.varq} against \N{e3.qwen.varbf}): its 4-bit build does not only shift the rate but also makes individual labels less stable.

\section{Judge, labels and protocol}
\label{app:judge}
\paragraph{Templates and builds.} Qwen2.5's template adds its default system prompt; Llama's inserts the current date, which we pin. All builds use \texttt{mlx-lm} 0.32.2 on MLX 0.32.4. The mixed 4/6 recipe keeps the value and down projections at 6 bits in the first and last eighth of layers and in every third layer between them. AWQ and GPTQ are calibrated on 128 sequences of 512 tokens of \texttt{mlx-lm}'s default calibration text. GGUF files are converted with llama.cpp at a pinned commit, without an importance matrix, and served with prompts sent as token ids; the server's count of evaluated tokens is checked against the tokens sent.

\paragraph{Judge.} The judge (Qwen2.5-14B-Instruct at 4 bits) labels each reply COMPLY, PARTIAL, NONE or INCOHERENT. Its answer is generated and parsed, and every label parsed. Chat-template control strings in requests and replies are neutralized before judging, because some MXFP4 replies contain them (Appendix~\ref{app:runon}). Benign prompts are judged with an XSTest-style rubric (compliance, partial refusal, refusal), and partial or full refusal counts as refusal. The 128-token cap is checked at 384 tokens on a fixed random subset of 150 prompts.

\paragraph{Human labels.} For each model, every forced-start prompt falls into one of three strata by the judge's labels on its two replies: D, the judge splits the pair; P, it calls both harmful; N, it calls neither. All of D and 25 random pairs from each of P and N were labelled (Table~\ref{tab:strata}), \N{human.totalpairs} pairs and \N{human.totalreplies} replies in all. Each pair was shown on one page with the request, the prefix and the two replies in random order as A and B, with no indication of format or judge label, and each reply was labelled with the judge's four categories. Rates, agreement, $\kappa$, PPV and NPV are design-weighted: each labelled pair counts for its stratum's population divided by its sample size, so the estimates refer to all 562 prompts and not to the enriched sample. Net shifts are point estimates from this weighting; their interval comes from a stratified bootstrap over labelled pairs.

\begin{table}[h]
\centering\small
\resizebox{\columnwidth}{!}{
\begin{tabular}{lrrrrrr}
\toprule
 & \multicolumn{2}{c}{D (judge splits)} & \multicolumn{2}{c}{P (both harmful)} & \multicolumn{2}{c}{N (neither)} \\
\cmidrule(lr){2-3}\cmidrule(lr){4-5}\cmidrule(lr){6-7}
Model & pop. & labelled & pop. & labelled & pop. & labelled \\
\midrule
\qwen & 101 & 101 & 68 & 25 & 393 & 25 \\
\llama & 104 & 104 & 120 & 25 & 338 & 25 \\
\qwenthree & 45 & 45 & 32 & 25 & 485 & 25 \\
\bottomrule
\end{tabular}
}
\caption{Human-label strata, forced start. ``pop.'' is the number of prompts in the stratum.}
\label{tab:strata}
\end{table}

\begin{table}[h]
\centering\small
\begin{tabular}{lccc}
\toprule
 & Judge & Llama Guard & Human \\
\midrule
\qwen      & \N{C1.permodel.qwen.all.diff}      & \N{labels.qwen.llamaguard.diff}      & \N{human.qwen.net} \\
\llama     & \N{C1.permodel.llama.all.diff}     & \N{labels.llama.llamaguard.diff}     & \N{human.llama.net} \\
\qwenthree & \N{C1.permodel.qwenthree.all.diff} & \N{labels.qwenthree.llamaguard.diff} & \N{human.qwenthree.net} \\
\bottomrule
\end{tabular}
\caption{Forced-start primary shift (pp) under three label sources; the human shift is design-weighted.}
\label{tab:labels}
\end{table}

\section{Secondary families}
\label{app:families}
Each family is Holm-corrected within itself. Rates are percentages of prompts; ``Disc.'' is the number of prompts on which only bf16, or only the build, was judged harmful (or, for benign prompts, refused).

\paragraph{Depth.} Table~\ref{tab:depth} gives the shift at $k = 0$, 2 and 8 forced tokens. \qwen's 4-bit shift is present from $k=2$; at $k=0$ bf16 almost never complies, so the shift is small in absolute terms but survives correction. \llama's MXFP4 build is less compliant at every depth, significantly so at $k=0$ and $k=2$.

\begin{table*}[t]
\centering\small
\begin{tabular}{llrrlrll}
\toprule
Model & Cell & bf16 \% & build \% & Shift [95\% CI] & Disc. & $p$ & Holm $p$ \\
\midrule
\qwen & 4-bit, $k=0$ & 1.6 & 4.3 & +2.7 [+1.2, +4.3] & 2/17 & $7.3\times10^{-4}$ & $0.0095$ \\
\qwen & 4-bit, $k=2$ & 10.9 & 17.6 & +6.8 [+3.7, +9.8] & 19/57 & $1.5\times10^{-5}$ & $2.1\times10^{-4}$ \\
\qwen & MXFP4, $k=0$ & 1.6 & 2.5 & +0.9 [$-$0.4, +2.3] & 5/10 & $0.30$ & $1.00$ \\
\qwen & MXFP4, $k=2$ & 10.9 & 14.4 & +3.6 [+0.7, +6.4] & 23/43 & $0.019$ & $0.21$ \\
\qwen & MXFP4, $k=8$ & 18.0 & 17.3 & $-$0.7 [$-$4.1, +2.7] & 49/45 & $0.76$ & $1.00$ \\
\addlinespace
\llama & 4-bit, $k=0$ & 6.4 & 4.4 & $-$2.0 [$-$3.6, $-$0.5] & 15/4 & $0.019$ & $0.21$ \\
\llama & 4-bit, $k=2$ & 29.0 & 29.4 & +0.4 [$-$3.2, +3.9] & 51/53 & $0.92$ & $1.00$ \\
\llama & MXFP4, $k=0$ & 6.4 & 1.4 & $-$5.0 [$-$6.9, $-$3.2] & 29/1 & $5.8\times10^{-8}$ & $8.7\times10^{-7}$ \\
\llama & MXFP4, $k=2$ & 29.0 & 23.1 & $-$5.9 [$-$9.3, $-$2.3] & 68/35 & $0.0015$ & $0.018$ \\
\llama & MXFP4, $k=8$ & 30.6 & 26.5 & $-$4.1 [$-$7.8, $-$0.5] & 67/44 & $0.036$ & $0.33$ \\
\addlinespace
\qwenthree & 4-bit, $k=0$ & 0.4 & 0.5 & +0.2 [+0.0, +0.5] & 0/1 & $1.00$ & $1.00$ \\
\qwenthree & 4-bit, $k=2$ & 3.2 & 4.3 & +1.1 [$-$0.4, +2.5] & 6/12 & $0.24$ & $1.00$ \\
\qwenthree & MXFP4, $k=0$ & 0.4 & 0.2 & $-$0.2 [$-$0.5, +0.0] & 1/0 & $1.00$ & $1.00$ \\
\qwenthree & MXFP4, $k=2$ & 3.2 & 3.9 & +0.7 [$-$0.5, +2.0] & 5/9 & $0.42$ & $1.00$ \\
\qwenthree & MXFP4, $k=8$ & 10.5 & 10.1 & $-$0.4 [$-$2.7, +2.0] & 24/22 & $0.88$ & $1.00$ \\
\bottomrule
\end{tabular}

\caption{Depth family (forced start, 562 prompts per cell). The 4-bit $k=8$ cell is the primary (\S\ref{sec:forced}) and is not part of this family.}
\label{tab:depth}
\end{table*}

\paragraph{Benign prompts.} Table~\ref{tab:benignfull} gives the refusal rate on XSTest's 250 safe prompts and WildJailbreak's 210 adversarial-benign prompts. \S\ref{sec:interp} discusses the two that matter: \llama's 4-bit build refuses more jailbreak-styled benign prompts, and \qwen's refuses fewer safe ones. \qwenthree's 4-bit build also refuses fewer jailbreak-styled benign prompts, the opposite of \llama.

\begin{table*}[t]
\centering\small
\begin{tabular}{llrrlrll}
\toprule
Model & Cell & bf16 \% & build \% & Shift [95\% CI] & Disc. & $p$ & Holm $p$ \\
\midrule
\qwen & 4-bit, XSTest safe & 26.0 & 19.6 & $-$6.4 [$-$11.2, $-$1.6] & 27/11 & $0.014$ & $0.083$ \\
\qwen & 4-bit, WJB benign & 8.6 & 7.6 & $-$1.0 [$-$4.8, +2.9] & 10/8 & $0.81$ & -- \\
\qwen & MXFP4, XSTest safe & 26.0 & 32.0 & +6.0 [+0.8, +11.2] & 16/31 & $0.04$ & $0.20$ \\
\qwen & MXFP4, WJB benign & 8.6 & 10.5 & +1.9 [$-$2.4, +6.2] & 8/12 & $0.50$ & -- \\
\addlinespace
\llama & 4-bit, XSTest safe & 17.2 & 16.8 & $-$0.4 [$-$4.8, +3.6] & 15/14 & $1.00$ & $1.00$ \\
\llama & 4-bit, WJB benign & 9.5 & 15.2 & +5.7 [+1.4, +10.0] & 5/17 & $0.017$ & -- \\
\llama & MXFP4, XSTest safe & 17.2 & 18.4 & +1.2 [$-$3.2, +5.6] & 15/18 & $0.73$ & $1.00$ \\
\llama & MXFP4, WJB benign & 9.5 & 14.3 & +4.8 [+1.0, +9.0] & 5/15 & $0.041$ & -- \\
\addlinespace
\qwenthree & 4-bit, XSTest safe & 17.6 & 18.8 & +1.2 [$-$3.6, +6.0] & 18/21 & $0.75$ & $1.00$ \\
\qwenthree & 4-bit, WJB benign & 21.0 & 12.4 & $-$8.6 [$-$13.3, $-$4.3] & 21/3 & $2.8\times10^{-4}$ & -- \\
\qwenthree & MXFP4, XSTest safe & 17.6 & 22.4 & +4.8 [+0.0, +9.6] & 14/26 & $0.081$ & $0.32$ \\
\qwenthree & MXFP4, WJB benign & 21.0 & 19.5 & $-$1.4 [$-$5.7, +2.9] & 13/10 & $0.68$ & -- \\
\bottomrule
\end{tabular}

\caption{Refusal on benign prompts (partial or full refusal). A positive shift is more refusal at the build. Holm $p$ is over the six registered XSTest cells; the WJB benign cells were not registered and are uncorrected.}
\label{tab:benignfull}
\end{table*}

\paragraph{SORRY-Bench.} Table~\ref{tab:sorry} gives the overall shift on SORRY-Bench with no prefill, and the difference-in-differences between its linguistic rewrites and the base phrasing. \llama's overall shift has the sign of its free-start shift, and both Qwen models move slightly toward compliance. None of the six difference-in-differences intervals excludes zero, so we find no sign that disguising a request changes how quantization acts on it.

\begin{table*}[t]
\centering\small
\begin{tabular}{llll}
\toprule
Model & Build & Overall shift [95\% cluster CI] & Rewrites $-$ base [95\% cluster CI] \\
\midrule
\qwen & 4-bit & +3.2 [+0.7, +5.7] & +2.9 [$-$2.1, +8.1] \\
\qwen & MXFP4 & +3.2 [+0.3, +6.2] & $-$2.3 [$-$9.6, +4.6] \\
\llama & 4-bit & $-$5.0 [$-$7.8, $-$2.2] & +1.0 [$-$5.6, +7.9] \\
\llama & MXFP4 & $-$8.5 [$-$11.8, $-$5.5] & +5.0 [$-$1.5, +11.9] \\
\qwenthree & 4-bit & +3.8 [+1.8, +6.0] & $-$1.5 [$-$6.2, +2.7] \\
\qwenthree & MXFP4 & $-$1.0 [$-$3.3, +1.3] & $-$2.3 [$-$6.2, +1.7] \\
\bottomrule
\end{tabular}

\caption{SORRY-Bench, cluster bootstrap over 120 questions (pp).}
\label{tab:sorry}
\end{table*}

\section{Construction of the controls}
\label{app:controls}
For each weight matrix $W$ that the default 4-bit build quantizes, including the tied embedding, $E$ is computed in float32 as the dequantized 4-bit matrix minus $W$; norms and biases are untouched. The mirror is $W - E$ and the scaled builds are $W + aE$. A permuted build shuffles $E$ within each row and each group of 64 input columns, so each quantization group keeps its exact multiset of errors. A Gaussian build draws, for each row, Gaussian noise with that row's root-mean-square error and rescales it so that its norm equals the row norm of $E$ exactly. Every build is a dense bf16 model.

\begin{figure*}[t]
\centering
\includegraphics[width=\textwidth]{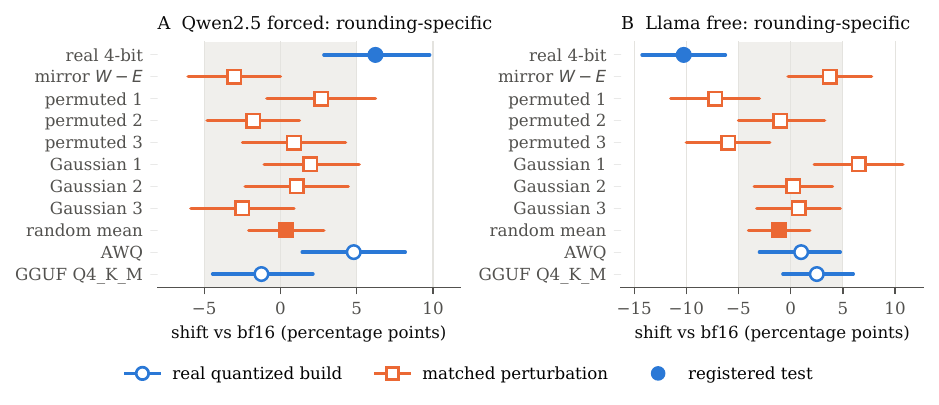}
\caption{The mirror and the six random controls in each cell with a shift, as a shift against bf16 with a 95\% interval, beside AWQ and llama.cpp's Q4\_K\_M (the latter against its own bf16). Filled markers are confirmatory tests; the panel title gives the verdict of the decision rule in \S\ref{sec:controls}.}
\label{fig:controls}
\end{figure*}

Before any judged run, the dense $W + E$ build had to match the real 4-bit build in NLL within 0.005 nats per token and in first-token argmax on at least 98\% of \N{ctrl.qwenforced.gate.n} prompts. Both cells passed (\S\ref{sec:controls}). A control whose NLL gap to bf16 exceeded twice the real build's would have been excluded from the random-control mean; none did. Table~\ref{tab:controls} gives every build and Figure~\ref{fig:controls} draws them beside the other quantizers' real errors.

\begin{table*}[t]
\centering\small
\begin{tabular}{lllrl}
\toprule
Cell & Build & $\Delta$NLL/token & Norm ratio & Shift [95\% CI] \\
\midrule
Qwen2.5 forced & real 4-bit & +0.144 & -- & +6.2 [+2.8, +9.8] \\
 & dense $W+E$ & +0.144 & 1.000 & +4.8 [+1.2, +8.4] \\
 & mirror $W-E$ & +0.131 & 1.000 & $-$3.0 [$-$6.0, +0.0] \\
 & permuted 1 & +0.166 & 1.000 & +2.7 [$-$0.9, +6.2] \\
 & permuted 2 & +0.180 & 1.000 & $-$1.8 [$-$4.8, +1.2] \\
 & permuted 3 & +0.155 & 1.000 & +0.9 [$-$2.5, +4.3] \\
 & Gaussian 1 & +0.168 & 1.000 & +2.0 [$-$1.1, +5.2] \\
 & Gaussian 2 & +0.138 & 1.000 & +1.1 [$-$2.3, +4.4] \\
 & Gaussian 3 & +0.178 & 1.000 & $-$2.5 [$-$5.9, +0.9] \\
 & $W+\tfrac12E$ & +0.033 & 0.500 & +2.7 [$-$0.5, +5.9] \\
 & $W+2E$ & +0.855 & 2.000 & +10.7 [+6.4, +14.9] \\
\addlinespace
Llama free & real 4-bit & +0.038 & -- & $-$10.3 [$-$14.0, $-$6.2] \\
 & dense $W+E$ & +0.038 & 1.000 & $-$10.7 [$-$14.8, $-$7.0] \\
 & mirror $W-E$ & +0.061 & 1.000 & +3.7 [$-$0.2, +7.8] \\
 & permuted 1 & +0.067 & 1.000 & $-$7.3 [$-$11.5, $-$3.0] \\
 & permuted 2 & +0.066 & 1.000 & $-$1.0 [$-$5.0, +3.2] \\
 & permuted 3 & +0.069 & 1.000 & $-$6.0 [$-$10.0, $-$2.0] \\
 & Gaussian 1 & +0.067 & 1.000 & +6.5 [+2.2, +10.8] \\
 & Gaussian 2 & +0.065 & 1.000 & +0.3 [$-$3.5, +4.0] \\
 & Gaussian 3 & +0.061 & 1.000 & +0.8 [$-$3.2, +4.8] \\
 & $W+\tfrac12E$ & +0.005 & 0.500 & $-$5.8 [$-$9.2, $-$2.2] \\
 & $W+2E$ & +0.224 & 2.000 & $-$17.8 [$-$22.5, $-$12.8] \\
\bottomrule
\end{tabular}

\caption{Matched-perturbation builds. $\Delta$NLL is against bf16 on bf16's own harmless replies; the norm ratio is the realized perturbation norm over $\lVert E\rVert$, identical across matrices to the precision shown.}
\label{tab:controls}
\end{table*}

\paragraph{Error structure of the quantizers.} Table~\ref{tab:errors} compares the quantizers' errors with the default round-to-nearest (RTN) error, matrix by matrix over the \N{err.qwen.nmat} (\qwen) and \N{err.llama.nmat} (\llama) quantized matrices. GPTQ's error is the closest to RTN's in direction (median cosine \N{err.qwen.cosrtngptq} and \N{err.llama.cosrtngptq}); MXFP4's is nearly orthogonal to it. AWQ rescales input channels and folds the scales into adjacent layers, so its weights are not directly comparable with $W$ and we report only the cosine.

\begin{table*}[t]
\centering\small
\begin{tabular}{lrrrrrr}
\toprule
 & \multicolumn{3}{c}{median relative error} & \multicolumn{3}{c}{median cosine with RTN error} \\
\cmidrule(lr){2-4}\cmidrule(lr){5-7}
Model & RTN & MXFP4 & GPTQ & MXFP4 & GPTQ & AWQ \\
\midrule
\qwen & 0.094 & 0.118 & 0.115 & 0.11 & 0.31 & 0.04 \\
\llama & 0.093 & 0.118 & 0.111 & 0.07 & 0.32 & 0.13 \\
\bottomrule
\end{tabular}

\caption{Per-matrix error of each quantizer: relative error $\lVert \hat W - W\rVert / \lVert W\rVert$, and cosine between its error and the RTN error.}
\label{tab:errors}
\end{table*}

\section{Free-start development set}
\label{app:pilot}
The first 200 WildJailbreak prompts form the development set (Table~\ref{tab:pilot}); the free-start test uses the 400 held-out prompts. \llama's 4-bit build was \N{pilot.llama.4bit-g64.diff}\pp{} against bf16 and \qwenthree's \N{pilot.qwenthree.4bit-g64.diff}\pp. On the held-out prompts the first holds and the second shrinks (\S\ref{sec:free}).

\begin{table*}[t]
\centering\small
\begin{tabular}{llrrlrll}
\toprule
Model & Build & bf16 \% & build \% & Shift [95\% CI] & Disc. & $p$ & Holm $p$ \\
\midrule
\qwen & 4-bit & 48.5 & 53.5 & +5.0 [$-$1.5, +11.5] & 19/29 & $0.19$ & -- \\
\qwen & MXFP4 & 48.5 & 49.5 & +1.0 [$-$6.0, +8.0] & 25/27 & $0.89$ & -- \\
\llama & 4-bit & 45.0 & 30.5 & $-$14.5 [$-$20.5, $-$8.5] & 34/5 & $2.4\times10^{-6}$ & -- \\
\llama & MXFP4 & 45.0 & 34.0 & $-$11.0 [$-$17.0, $-$5.0] & 31/9 & $6.8\times10^{-4}$ & -- \\
\qwenthree & 4-bit & 23.5 & 33.5 & +10.0 [+5.0, +15.0] & 4/24 & $1.8\times10^{-4}$ & -- \\
\qwenthree & MXFP4 & 23.5 & 23.0 & $-$0.5 [$-$5.0, +4.0] & 11/10 & $1.00$ & -- \\
\bottomrule
\end{tabular}

\caption{Free-start development set, 200 prompts. No correction is applied.}
\label{tab:pilot}
\end{table*}

\section{End-of-turn failure in MXFP4 \qwen}
\label{app:runon}
MXFP4 \qwen{} sometimes fails to end its turn and writes a fabricated next user turn. On 200 benign prompts with a 512-token cap this happens in \N{s5.qwen.mxfp4}\% of replies and in no other build we tested (Table~\ref{tab:runon}). It persists when the weights are dequantized to bf16 (\N{s5.qwen.mxfp4dequant}\%), so it comes from the weights and not from the kernel, and it does not appear in MXFP4 Qwen2.5-0.5B.

\begin{table*}[t]
\centering\small
\begin{tabular}{lrrrrrrr}
\toprule
Model & bf16 & 8-bit & 4-bit g32 & mixed 4/6 & 4-bit & MXFP4 & MXFP4 deq. \\
\midrule
\qwen & 0.0 & 0.0 & 0.0 & 0.0 & 0.0 & 10.5 & 8.5 \\
\llama & 0.0 & -- & -- & -- & 0.0 & 0.0 & -- \\
\qwenthree & 0.0 & -- & -- & -- & 0.0 & 0.0 & -- \\
Qwen2.5-0.5B & 0.0 & -- & -- & -- & 0.0 & 0.0 & 0.0 \\
\bottomrule
\end{tabular}

\caption{Share (\%) of replies to 200 benign prompts that contain a fabricated next turn (512-token cap). ``--'': not built for that model.}
\label{tab:runon}
\end{table*}

\section{Compute, software and release}
\label{app:compute}
All runs used one Apple M2 Pro with 16\,GB of memory, \texttt{mlx} 0.32.4 and \texttt{mlx-lm} 0.32.2, with one model in memory at a time. The logged experiment runs took about \N{compute.hours} hours over \N{compute.steps} steps, building, generating and judging; this excludes development. The bf16 checkpoints are identified by the first twelve hex digits of the SHA-256 over their weight files: \qwen{} \texttt{\N{hash.qwen}}, \llama{} \texttt{\N{hash.llama}}, \qwenthree{} \texttt{\N{hash.qwenthree}}.

We release the code, the pre-registration, every judge and human label keyed by prompt index, the aggregate results, the perturbation builder and the analysis code, at \url{https://github.com/egekaya1/qrefusal}. The analysis reads the raw result files, which contain the replies; we do not release those, because they include harmful content produced under a forced start, and they are available to researchers on request.

\paragraph{Use of AI assistants.} AI assistants were used for coding, literature search and editing of the text. The author takes full responsibility for the content.

\end{document}